\documentclass[1p]{elsarticle}
\usepackage{ifpdf}
\usepackage{graphicx}
\usepackage{amssymb}
\usepackage{amsmath}
\biboptions{sort&compress}
\journal{Nonlinear Analysis: Real World Applications}

\begin{document}

\begin{frontmatter}

\title{A note on deriving linearizing transformations for a class of second order nonlinear ordinary differential equations}

\author[mymainaddress]{R. Mohanasubha}

\author[mysecondaryaddress]{V. K. Chandrasekar}
%\ead{}

\author[mymainaddress]{M. Senthilvelan\corref{mycorrespondingauthor}}
\cortext[mycorrespondingauthor]{Corresponding author}
\ead{velan@cnld.bdu.ac.in}

\address[mymainaddress]{Centre for Nonlinear Dynamics, School of Physics, Bharathidasan University, Tiruchirappalli-620024, Tamil Nadu, India}
\address[mysecondaryaddress]{Centre for Nonlinear Science and Engineering, School of Electrical and Electronics Engineering, SASTRA University, Thanjavur-613401, Tamil Nadu, India}

\begin{abstract}
We present a method of deriving linearizing transformations for a class of second order nonlinear ordinary differential equations.  We construct a general form of a nonlinear ordinary differential equation that admits Bernoulli equation as its first integral.  We extract conditions for this integral to yield three different linearizing transformations, namely point, Sundman and generalized linearizing transformations.  The explicit forms of these linearizing transformations are given.  The exact forms and the general solution of the nonlinear ODE for these three linearizables cases are also enumerated.  We illustrate the procedure with three different examples. 
\end{abstract}

\begin{keyword}
Linearizing transformations \sep  Ordinary differential equations \sep General solutions
\end{keyword}

\end{frontmatter}

%\linenumbers

\section{Introduction} 

Nonlinear ordinary differential equations (ODEs) can be solved in a number of ways. For example, integration by quadrature, exploring Darboux polynomials and/or Jacobi last multipliers, order reduction procedure through symmetry methods, direct linearization and so on 
\cite{zait,book1,ste,olv,ibra,nuci2,bluman}. Among the above, the linearization of nonlinear ODEs got impetus in recent years \cite{ anna1,recent1, recent2, recent3, recent4, ffl, mur_new1}. The task here is to transform the given nonlinear ODE into a linear ODE whose solution is known. Confining our attention on second order nonlinear ODEs a systematic study on this problem has been initiated by Sophus Lie long ago 
\cite{olv}. He demonstrated that the nonlinear ODEs that can be transformed to free particle equation should be cubic in first derivative whose coefficients (which are functions of $t$ and $x$) should satisfy a couple of second order partial differential equations \cite{olv, ibra}. 
The underlying linearizing transformation (LT) is considered as point transformation (PT) since the new dependent $(w)$ and independent $(z)$ variables are functions of old dependent $(x)$ and independent $(t)$ variables alone, that is $w=F(t,x)$ and $z=G(t,x)$ \cite{ibra}. Later it has been shown that the linearizable ODEs under point transformations admit maximal Lie point symmetries \cite{ibra}. Subsequently linearizing PTs have been identified from the Lie point symmetries \cite{olv} itself.  

The linearization of nonlinear ODEs under nonlocal transformation has also been investigated in detail \cite{berkov,dua, moyo, euler}. The necessary and sufficient condition for a second order nonlinear ODE to be linearizable under the Sundman transformation (ST), $w=f(t,x)$, $z=\int g(t,x)dt$, had been analyzed by Duarte et al. \cite{dua}. Here $f$ and $g$ are functions of $t$ and $x$ only. Unlike the PT the new independent variable $z$ is considered in nonlocal form. Since the independent variable $z$ is nonlocal, it is difficult to invert the solution from its linear counterpart \cite{ibra}. The connection between $\lambda$-symmetries and Sundman linearizable ODEs of second-order ODEs has been analyzed by few authors, see for example Refs.\cite{mur1,mur2} and references therein.
 
Apart from the above two LTs, second-order nonlinear ODEs can also be linearized by generalized linearizing transformations (GLT), namely $w=f(t,x)$ and $z=\int g(t,x,\dot{x})dt$ \cite{anna1,chanuni}. The difference between ST and GLT is that in the latter the new independent variable $z$ is generalized to contain the first derivative. Unlike the above two LTs the connection between GLT and symmetries is not known. However, it has been demonstrated that equations which cannot be linearized by PT and ST can be linearized by GLT \cite{chanuni}.

It is clear from the above facts that in linearization besides identifying the LTs one has to device a suitable procedure to derive the general solution of the considered nonlinear ODE. In this context, recently a simple but powerful method of identifying the LTs has been introduced \cite{anna1}. In this method the LTs have been identified from the first integral itself by rewriting it as a ratio of two perfect derivative functions. A perfect derivative function that appears in the numerator acts as the new dependent variable $(w)$ and the function that appears as a perfect derivative in the denominator acts as the new independent variable $(z)$,  which inturn linearizes the nonlinear ODE. Interestingly, unlike the other methods, this procedure readily gives all the aforementioned LTs, namely PT, ST and GLT in a simple and straightforward manner \cite{anna1}. This method also produced several new LTs in higher order ODEs which are previously unknown \cite{anna1}.

In the earlier works, the usefulness of this method has been demonstrated only for specific examples. In this work, we derive a general form of second order nonlinear ODE that can be linearized by the aforementioned LTs. We also derive the first integral for the considered nonlinear ODE.  We then extract the conditions for this integral to yield PT.  We also present the explicit form of the PT that linearizes the nonlinear ODE. We then move on to identify the condition for this integral to give the ST. The explicit form of the ST that linearizes the nonlinear ODE is also given. Finally, by rewriting the integral suitably we explore the explicit form of the GLT that linearizes yet another family of ODEs.  Our result shows that in the case of Sundman linearizable, the first integral should be a linear polynomial in $\dot{x}$. This result agrees with the one reported earlier \cite{mur1}. In other words once a nonlinear ODE is identified as a linearizable equation then one can follow the procedure given in this paper and identify the LT readily. In the earlier works, the general solution of the nonlinear ODE has been derived only from the known solution of its linear counterpart which pose some obstacles in the case of ST and GLT since the independent variable in them are nonlocal. Differing from that, in this work, the general solution of the nonlinear ODE is given explicitly for all three cases. The general solution is derived by integrating the integral directly. The procedure developed in this paper not only gives the LTs but also its general solution in a straightforward way.

The structure of the paper is as follows: In section 2, we identify the general form of the nonlinear ODE that admits Bernoulli equation as its first integral. In section 3, we present the method of obtaining LTs from the first integral. We also derive the conditions to obtain PT and ST from this integral. The explicit forms of these two LTs are given.  We then extract the explicit form of the GLT from the integral. In section 4, we consider three different examples, one for PT, second for ST and last one for GLT, and demonstrate the method of identifying the LTs, first integral and the general solution. Finally, in Sec. 5, we present our conclusion.

\section{Explicit form of the linearizable equation}
As our aim to derive the LTs from the first integral we start our analysis by considering an integral which is rational in $\dot{x}$, that is
\begin{equation}
I(t,x,\dot{x})=\frac {A(t,x)\dot{x}+B(t,x)} {C(t,x)\dot{x}+D(t,x)}\label{e2},
\end{equation}
where the functions $A, B, C$ and $D$ are functions of $x$ and $t$ whose exact expressions are to be determined. 

The integral has been assumed in a natural way. For a class of nonlinear ODEs the first integral can be written either in a polynomial form or in the rational form. For example, the nonlinear oscillator equation $\ddot{x}-\frac{2}{x}\dot{x}^2+\frac{2x}{t^2}=0$ admits the first integral in the form $\frac{t\dot{x}-x}{t^2x^2}$ whereas the nonlinear oscillator equation $\ddot{x}-\frac{3x\dot{x}^2}{x}-\frac{\dot{x}}{t}=0$ admits the first integral in the form $\frac{\dot{x}}{tx^3}$. In fact a class of nonlinear oscillator equations of the form $\ddot{x}+f(x)\dot{x}^2+g(x)\dot{x}+h(x)=0$ admit the first integral in the rational form.  With this observation, in our analysis, we consider the integral in the form (\ref{e2}).

To determine the nonlinear ODE that admits (\ref{e2}) as its first integral, we proceed as follows. We rewrite the first integral (\ref{e2}) to obtain 
\begin{equation}
\dot{x}=\frac {\hat{B}-I\hat{D}} {\hat{C}I-1}.\label{32m}
\end{equation}
where $\frac{N} {A}=\hat{N}$, $N = B, C, D$. For simplicity, we consider $\hat{C}(t,x)$ is a function of $t$ alone, that is $\hat{C}=f(t)$. Now defining the denominator as a new function, say $\tilde{f}(t)=\frac{1}{f(t)I-1}$, Eq.(\ref{32m}) can be rewritten in the form
\begin{equation}
\dot{x}=\tilde{f}(t)(\hat{B}-I\hat{D}).\label{33m}
\end{equation}
Since linearizable nonlinear ODEs are solvable, the corresponding order reduced nonlinear ODE  (\ref{33m}) not only be solvable but its solution should also be known. The most general first order nonlinear ODE whose solution  explicitly known is the Bernoulli equation. Comparing the first order ODE (\ref{33m}) with Bernoulli equation, we can fix the functions, $\hat{B}$ and $\hat{D}$, are of the form
\begin{eqnarray}
\hat{B}=r_{1}(t)x+ r_{2}(t)x^q,~~
\hat{D}=r_{3}(t)x+r_{4}(t)x^q,
\end{eqnarray}
where $r_i(t),~i=1,2,3,4,$ are functions of $t$ and $q$ is an integer. Substituting these forms in (\ref{33m}), we obtain
\begin{equation}
 \dot{x}=a(t)x+b(t)x^{q},\label{beer}
\end{equation}
where
\begin{eqnarray}
a(t)=\tilde{f}(t)[(r_{1}(t)-Ir_{3}(t))],~~
b(t)=\tilde{f}(t)[(r_{2}(t)-Ir_{4}(t))].\label{37m}
\end{eqnarray}
The Bernoulli equation (\ref{beer}) can be derived from the integral %(\ref{e2})as
\begin{equation}
I(t,x,\dot{x})=\frac {\dot{x}+r_{1}(t)x+r_{2}(t)x^{q}} {\dot{x}f(t)+r_{3}(t)x+r_{4}(t)x^{q}}\label{k31_new}.
\end{equation}
The second order nonlinear ODE which admits the first integral (\ref{k31_new}) should be of the form
\begin{eqnarray}
&\displaystyle{\ddot{x}+a_{2}(t,x)\dot{x}^{2}+a_{1}(t,x)\dot{x}+a_{0}(t,x)=0 }.\label{main2}
\end{eqnarray}
The coefficients $a_2,~a_1$ and $a_0$ can be expressed in terms of the functions $r_i,~i=1,2,3,4$ and $f(t)$, that is 
\scriptsize
\begin{subequations}
\label{coefficient}
\begin{eqnarray} 
\hspace{-2cm}a_{2}(t,x)&=&\frac {r_{1}(t)f(t)-\dot{f}(t)-r_{3}(t)+(f(t)r_{2}(t)-r_{4}(t))qx^{q-1}} {\Lambda}, \label{a2_coeff} \\ 
\hspace{-2cm}a_{1}(t,x)&=&\frac {(f(t)\dot{r_{2}}(t)+r_{3}(t)r_{2}(t)q+r_{1}(t)r_{4}(t)-r_{1}(t)r_{4}(t)q-r_{2}(t)r_{3}(t)-r_{2}(t)\dot{f}(t)-\dot{r_{4}}(t))x^q} {\Lambda}\nonumber \\
&&+\frac{(\dot{r_{1}}(t)f(t)-r_{1}\dot{f}(t)-\dot{r_{3}}(t))x} {\Lambda}, \label{a1_coeff}\\ 
\hspace{-2cm}a_{0}(t,x)&=&\frac {(\dot{r_{1}}(t)r_{3}(t)-r_{1}(t)\dot{r_{3}}(t))x^2+(r_{3}(t)\dot{r_{2}}(t)+r_{4}(t)\dot{r_{1}}(t)-r_{1}(t)\dot{r_{4}}(t)-r_{2}(t)\dot{r_{3}}(t))x^{q+1}}  {\Lambda}\nonumber \\
 &&+\frac{(r_{4}(t)\dot{r_{2}}(t)-r_{2}(t)\dot{r_{4}}(t))x^{2q}}  {\Lambda},\label{jkd}
\end{eqnarray}
\end{subequations}
\normalsize
with $\Lambda=(r_{3}(t)-r_{1}(t)f(t))x+(r_{4}(t)-r_{2}(t)f(t))x^q$. 

In the following, we discuss the method of identifying the LTs from the first integral.

\section{ Method of obtaining the LTs} 
\subsection{General Theory}
Let us assume that the second order nonlinear ODE (\ref{main2}) admits a first integral $I(t,x, \dot{x})$ which is constant on the solutions. Now rewriting the first integral $I=f(t,x,\dot{x})$ as a product of two terms, namely a perfect derivative $\frac{dF}{dt}$ and $\frac{1}{G(t,x,\dot{x})}$, that is
\begin{equation}
I=\frac{1} {G(t,x,\dot{x})}\frac{d} {dt}F(t,x)\label{k1}.
\end{equation}
Suppose the function $G(t,x,\dot{x})$ is also a perfect derivative of another function $z$, that is $G=\frac{dz(t,x)} {dt}$, then the above first integral can be simplified to
\begin{equation}
I=\frac{\frac{dF}{dt}}{\frac{dG}{dt}}=\frac{dF}{dG}=\frac{dw} {dz},\label{fgrr}
\end{equation}
where $w$ and $z$ are new dependent and independent variables, respectively. Differentiating Eq. (\ref{fgrr}) with respect to $z$, we find
\begin{equation}
\frac{d^2w}{dz^2}=0.
\label{eq:1}
\end{equation}
In the above, 
\begin{equation}
w=F(t,x)~ \mathrm{and}~ z=\int G(t,x,\dot{x}) dt.
\label{w}
\end{equation}
Solving Eq. (\ref{eq:1}) we find $w=I_1 z+I_2$, where $I_1$ and $I_2$ are integration constants. From the latter expression we can deduce the solution of the nonlinear ODE. The new variables $w$ and $z$ are nothing but the LT for the given second order nonlinear ODE since they transform the given second order nonlinear ODE into the free particle equation (\ref{eq:1}).

The nature of $\omega$ and $z$ fixes the LTs to be a PT or ST or GLT. For example, if the function $G$ in (\ref{k1}) is independent of the variable 
$\dot{x}$ then it becomes an ST. On the other hand $G$ is a perfect 
differentiable function then it becomes an PT, that is,
$G(t,x,\dot{x})=\frac{d}{dt}\hat{G}(t,x)$, then
$dT=\frac{d\hat{G}}{dt}dt\Rightarrow T=\hat{G}(t,x)$. The transformation given in (\ref{w}) is nothing but the GLT. 

\subsection{Identifying linearizing transformations of (\ref{k31_new})}
To identify the LTs from the integral (\ref{k31_new}) we have to rewrite the numerator and denominator of it as two separate perfect differentiable functions. To begin, let us concentrate on the numerator. To write this function, $(\dot{x}+r_1 x+r_2x^q)$, as a perfect derivative we multiply it by a function $R_1$ (with the assumption that $R_1$ has to be determined) so that it becomes $R_1(\dot{x}+r_1 x+r_2 x^q)=\frac{dF}{dt}=F_t+\dot{x}F_x$. Now comparing the coefficient of $\dot{x}$ on both sides we find $F_x=R_1$ and $F_t=R_1(r_1 x+r_2 x^q)$. Integrating this system of equations, we find 
\begin{equation}
R_{1}=(1-q)x^{-q}e^{(1-q)\int{r_{1}(t)dt}}\label{k6}.
\end{equation}
Following the same procedure on the denominator we observe that it can be rewritten as a perfect derivative upon multiplying by the function
\begin{equation}
R_{2}=\frac{(1-q)x^{-q}e^{(1-q)\int{\frac{r_{3}} {f(t)}dt}}} {f(t)}\label{k8}.
\end{equation}
With the help of $R_1$ and $R_2$ the first integral (\ref{k31_new}) can now be brought to the form

\begin{equation}
I=\frac{\frac{1} {(1-q)x^{-q}e^{(1-q)\int{r_{1}(t)dt}}}\frac{d} {dt}\big(x^{(1-q)}e^{(1-q)\int{r_{1}(t)dt}}+(1-q)\int{e^{(1-q)\int{r_{1}(t)dt}}r_{2}(t)dt}\big)} {\frac{f(t)} {(1-q)x^{-q}e^{(1-q)\int{\frac{r_{3}} {f(t)}dt}}}\frac{d} {dt}\big(x^{(1-q)}e^{(1-q)\int{\frac{r_{3}} {f(t)}dt}}+(1-q)\int{e^{(1-q)\int{\frac{r_{3}} {f(t)}dt}}\frac{r_{4}} {f(t)}dt}\big)}\label{k9}.
\end{equation}
It is straightforward to verify that upon evaluating the total derivatives that appear in (\ref{k9}) and multiplying by their prefactors and simplifying the resultant expression it reduces to (\ref{k31_new}). In other words, the integral (\ref{k31_new}) has been just rewritten as ratio of two perfect derivative functions with some prefactors. Since the integral (\ref{k31_new}) has now been rewritten in the desired form, we can identify the LT from it in a straightforward manner. 
\subsection{Point transformation}
 The PT can be extracted from the first integral (\ref{k9}) provided the prefactors that appear in the numerator and denominator should cancel each other. Upon imposing this constraint, we find  
\begin{equation}
r_{1}(t)=\frac{\dot{f}(t)} {f(t)(q-1)}+\frac{r_{3}(t)} {f(t)}\label{k10}.
\end{equation}
The second order nonlinear ODE (\ref{main2}) admits PT only if it satisfies the condition (\ref{k10}). With this constraint, the first integral now becomes
\begin{equation}
I=\frac{\frac{d} {dt}\big(x^{(1-q)}e^{(1-q)\int{r_{1}(t)dt}}+(1-q)\int{e^{(1-q)\int{r_{1}(t)dt}}r_{2}(t)dt}\big)} {\frac{d} {dt}\big(x^{(1-q)}e^{(1-q)\int{\frac{r_{3}} {f(t)}dt}}+(1-q)\int{e^{(1-q)\int{\frac{r_{3}} {f(t)}dt}}\frac{r_{4}} {f(t)}dt}\big)}\label{k11}.
\end{equation}
From (\ref{k11}) we can readily identify the linearizing PT and it is of the form
\begin{eqnarray}\nonumber
\omega&=&x^{(1-q)}e^{(1-q)\int{r_{1}(t)dt}}+(1-q)\int{e^{(1-q)\int{r_{1}(t)dt}}r_{2}(t)dt}\\
z&=&x^{(1-q)}e^{(1-q)\int{\frac{r_{3}(t)}{f(t)}dt}}+(1-q)\int{e^{(1-q)\int{\frac{r_{3}(t)} {f(t)}dt}}\frac{r_{4}(t)}{f(t)}dt}\label{k22}.
\end{eqnarray}

 We note here that for the given nonlinear ODE (\ref{main2}), $r_1,~r_2,~r_3$ and $r_4$ are known. Substituting their explicit forms in (\ref{k11}) and (\ref{k22}) one can get the first integral and the LT explicitly in a straightforward manner.
The general solution of (\ref{main2}) can be identified by integrating the Eq. (\ref{beer}). Upon imposing the constraint (\ref{k10}) in the general solution of (\ref{beer}), we find 
\begin{align}
x(t)^{1-q}=\bigg(Ce^{(1-q)\mu}+(1-q)e^{(1-q)\mu}\int e^{(q-1)\mu}\tilde{f}(t)[r_{2}(t)-Ir_{4}(t)]dt\bigg),\label{jnm_point}
\end{align}
where $\mu=\int \tilde{f}(t)[\frac{\dot{f}(t)} {f(t)(q-1)}+\frac{r_{3}(t)} {f(t)}-Ir_{3}(t)]dt$ and $C$ and $I$ are the two arbitrary constants. 
\subsection{Sundman transformation}
If the given equation does not satisfy the constraint (\ref{k10}), then the integral (\ref{k9}) provides either a ST or a GLT. As far as the ST is concerned, the new independent variable should not contain the derivative term inside the integral. With this in mind now let us rewrite the first integral (\ref{k9}) as
\begin{equation}
I=\frac{\frac{d} {dt}\bigg(x^{1-q}e^{(1-q)\int{r_{1}(t)dt}}+(1-q)\int{e^{(1-q)\int{r_{1}(t)}}r_{2}(t)dt}\bigg)} {\frac{f(t)e^{(1-q)\int{r_{1}(t)dt}}}{e^{(1-q)\int{\frac{r_{3}(t)} {f}dt}}}\frac{d}{dt}\bigg(x^{(1-q)}e^{(1-q)\int{\frac{r_{3}} {f(t)}dt}}+(1-q)\int{e^{(1-q)\int{\frac{r_{3}} {f(t)}dt}}\frac{r_{4}} {f(t)}dt}\bigg)}\label{k515}. 
\end{equation}
Let us focus our attention on the denominator. Evaluating the differentiation and multiplying by the prefactor given in (\ref{k515}), the denominator simplifies to
\begin{equation}
(1-q)e^{(1-q)\int{r_{1}(t)dt}}(fx^{-q}\dot{x}+x^{1-q}r_3(t)+r_4(t)).
\label{asa}
\end{equation}
This expression can be written as a perfect derivative in two different ways. One without $\dot{x}$ term inside the perfect derivative and the another with $\dot{x}$ term inside the perfect derivative. The first choice is possible only in the case $f(t)=0$. With this restriction, Eq. (\ref{asa}) can be written as
\begin{equation}
\frac{d}{dt}\bigg((1-q)\int e^{(1-q)\int{r_{1}(t)dt}}(r_3(t)x^{1-q}+r_4(t))\bigg)dt.
\label{eq:den}
\end{equation}
The integral (\ref{k515}) now reads
\begin{equation}
I=\frac{\frac{d} {dt}\bigg(x^{1-q}e^{(1-q)\int{r_{1}(t)dt}}+(1-q)\int{e^{(1-q)\int{r_{1}(t)}}r_{2}(t)dt}\bigg)} {\frac{d}{dt}\bigg((1-q)\int{e^{(1-q)\int{r_{1}(t)dt}}\bigg(r_{3}(t)x^{1-q}+r_{4}(t)\bigg)dt}\bigg)}\label{k55}.
\end{equation}
From (\ref{k55}) we can identify the LT as
\begin{eqnarray}\nonumber
\omega&=&x^{1-q}e^{(1-q)\int{r_{1}(t)dt}}+(1-q)\int{e^{(1-q)\int{r_{1}(t)dt}}r_{2}(t)dt},\\ 
z&=&(1-q)\int{e^{(1-q)\int{r_{1}(t)dt}}(r_{3}(t)x^{1-q}+r_{4}(t))dt}\label{k66}.
\end{eqnarray}
From (\ref{k66}) it is clear that the new independent variable $z$ has been identified in nonlocal form and it is a function of $t$ and $x$ alone. Suppose the given equation is Sundman linearizable then substituting the expressions $r_1,~r_2,~r_3$ and $r_4$ in (\ref{k66}) one can readily obtain the ST. Since $f(t)=0$, the integral (\ref{k31_new}) now becomes 
\begin{equation}
I(t,x,\dot{x})=\frac {\dot{x}+r_{1}(t)x+r_{2}(t)x^{q}} {r_{3}(t)x+r_{4}(t)x^{q}}=P(t,x)\dot{x}+Q(t,x),\label{inte_sund}
\end{equation}
where
\begin{equation}
P(t,x)=\frac{1}{r_{3}(t)x+r_{4}(t)x^{q}},~~Q(t,x)=\frac{r_{1}(t)x+r_{2}(t)x^{q}} {r_{3}(t)x+r_{4}(t)x^{q}}.
\end{equation}
As far as ST is concerned we have an integral which is a polynomial in $\dot{x}$. This result agrees with the one reported earlier \cite{mur1}.

The general solution of the nonlinear ODE (\ref{main2}) can be obtained by integrating (\ref{inte_sund}). Doing so, we find 
\begin{equation}
x(t)^{1-q}=e^{(1-q)\int(Ir_3(t)-r_1(t))dt}(C+(1-q)\int e^{(q-1)\int(Ir_3(t)-r_1(t))dt} (Ir_4(t)-r_2(t)) dt),\label{sol_sund}
\end{equation}
where $C$ and $I$ are the integration constants.

\subsection{Generalized linearizing transformation}
In case $f(t) \neq 0$ for the given nonlinear ODE then the denominator can be rewritten as a perfect derivative only in the form 
\begin{equation}
\frac{d}{dt} \bigg((1-q)\int{e^{(1-q)\int{r_{1}(t)dt}}\bigg(x^{-q}f(t)\dot{x}+r_{3}(t)x^{1-q}+r_{4}(t)\bigg)dt}\bigg) 
\end{equation}
so that the integral now becomes
\begin{equation}
I=\frac{\frac{d} {dt}\bigg(x^{1-q}e^{(1-q)\int{r_{1}(t)dt}}+(1-q)\int{e^{(1-q)\int{r_{1}(t)}}r_{2}(t)dt}\bigg)} {\frac{d}{dt}\bigg((1-q)\int{e^{(1-q)\int{r_{1}(t)dt}}\bigg(x^{-q}f(t)\dot{x}+r_{3}(t)x^{1-q}+r_{4}(t)\bigg)dt}\bigg)}\label{nkm}.
\end{equation}
From (\ref{nkm}) we can readily identify the LT as 
\begin{eqnarray}\nonumber
\omega&=&x^{1-q}e^{(1-q)\int{r_{1}(t)dt}}+(1-q)\int{e^{(1-q)\int{r_{1}(t)dt}}r_{2}(t)dt},\\ 
z&=&(1-q)\int{e^{(1-q)\int{r_{1}(t)dt}}(x^{-q}f(t)\dot{x}+r_{3}(t)x^{1-q}+r_{4}(t))dt}\label{nkm1}.
\end{eqnarray}
Since the new independent variable does admit the variable $\dot{x}$ we call this transformation as GLT.

 The general solution of (\ref{main2}) that can be linearized by GLT is given by %can be readily given in the form
\begin{align}
x(t)^{1-q}=\bigg(Ce^{(1-q)\mu}+(1-q)e^{(1-q)\mu}\int e^{(q-1)\mu}\tilde{f}(t)[r_{2}(t)-Ir_{4}(t)]dt\bigg),\label{jnm_gene}
\end{align}
where $\mu=\int \tilde{f}(t)[r_{1}(t)-Ir_{3}(t)]dt$ and $C$ and $I$ are the two arbitrary constants.

\section{Examples}
In this section, we demonstrate the above procedure with three examples. In the first example we consider an ODE that is linearizable by PT and in the second example we consider an ODE that is linearizable by ST. Third example is included to illustrate the linearization through GLT. 
\subsection{Example 1: Point transformation}
Let us consider the following example
\begin{equation}
\ddot{x}+3K(t)x\dot{x}+K(t)^2x^3+\dot{K}(t)x^2+\lambda(t)x=0,\label{exam1}
\end{equation}
where $K(t)$ and $\lambda(t)$ are functions of $t$. In the sub-case, $K(t)=constant=K$ and $\lambda(t)=constant=\lambda$, Eq. (\ref{exam1}) becomes the well known modified Emden equation with linear external forcing whose solvability and dynamics (both classical and quantum) have been investigated extensively by various authors \cite{pre,chithi,nuc_new,parth}. 

 Equation (\ref{exam1}) satisfies the Lie's linearizibility criteria and hence it is linearizible by PT. It is tedious to determine the linearizing PT from the Lie symmetries \cite{bluman,olv}. In the following, the LTs and the solution are derived in a simple and straightforward manner through the above said procedure. To begin, let us construct the first integral of (\ref{exam1}).

 By equating the coefficients of (\ref{exam1}) with (\ref{coefficient}), we can find the first integral of the given equation. Since (\ref{exam1}) does not contain $\dot{x}^2$, we have $a_2=0$. Solving $a_2=0$ with (\ref{a2_coeff}), we obtain the following expressions
\begin{equation}
r_4(t)=r_2(t)f(t),~~r_3(t)=r_1(t)f(t)-\dot{f},\label{dfr1}
\end{equation}
where over dot represents the derivative with respect to $t$.  Substituting the above forms in (\ref{a1_coeff}) with $a_1=3 K(t)x$ and solving the resultant expressions we find
\begin{equation}
q=2,~~r_1(t)=\frac{\ddot{f}}{2\dot{f}},~~r_2(t)=K(t).
\end{equation} 
Substituting the above parameters in (\ref{jkd}) and equating the resultant expression with the coefficient $a_0=K(t)^2x^3+\dot{K}(t)x^2+\lambda(t)x$ and solving the underlying equations we obtain
\begin{eqnarray}
\lambda=\frac {\dddot{f}} {2\dot{f}}-\frac {3} {4}\frac {\ddot{f}^2} {\dot{f}^2}.\label{eq1_para}
\end{eqnarray}
With these parameters (\ref{dfr1})-(\ref{eq1_para}), the first integral of (\ref{exam1}) now reads (see Eq. (\ref{k31_new}))%becomes %$I$ is of the 
\begin{equation}
I=\frac{\dot{x}+\frac{\ddot{f}}{2\dot{f}}x+K(t)x^2} {f\dot{x}+(\frac{\ddot{f}f} {2\dot{f}}-\dot{f})x+K(t)fx^2}.
\end{equation}
Upon substituting (\ref{dfr1}) in (\ref{k10}) the latter constraint is satisfied. This in turn confirms that the given equation is linearizable.  The integral now reads 
\begin{equation}
I=\frac{\frac{d} {dt}\bigg(\frac{1} {x\dot{f}^{\frac{1} {2}}}-\int{\frac{1}{\dot{f}^{\frac{1}{2}}}K(t)dt}\bigg)} {\frac{d}{dt}\bigg(\frac{f} {x\dot{f}^{\frac{1} {2}}}-\int{\frac{f}{\dot{f}^{\frac{1}{2}}}K(t)dt}\bigg)}=\frac{\frac{dw}{dt}}{\frac{dz}{dt}}=\frac{dw}{dz}. \label{II}
\end{equation}
From (\ref{II}) we can readily identify the linearizing PT in the form
\begin{equation}
w=\frac{1} {x\dot{f}^{\frac{1} {2}}}-\int{\frac{1}{\dot{f}K(t)}^{\frac{1}{2}}dt},~~z=\frac{f} {x\dot{f}^{\frac{1} {2}}}-\int{\frac{f}{\dot{f}^{\frac{1}{2}}}K(t)dt}.
\end{equation}

 By substituting the parameters (\ref{dfr1}) - (\ref{eq1_para}) in (\ref{jnm_point}), we can get the general solution of (\ref{exam1}) as
\begin{equation}
x(t)=\frac{ e^{-\delta}}{(C-\int(e^{\delta}\tilde{f}K(t)(1-If))dt)},
\end{equation}
where $\delta=\int{\tilde{f}(\frac{\dddot{f}}{2\dot{f}}-I(\frac{\ddot{f}f}{2\dot{f}}-\dot{f}))dt}$ and  $C$ and $I$ are arbitrary constants.
\subsection{Example 2: Sundman transformation}
Let us consider the following ODE, namely
\begin{equation}
\ddot{x}-\frac{1}{x}\dot{x}^2-\frac{\dot{x}}{t}=0.\label{exam2}
\end{equation}
Eq. (\ref{exam2}) passes the Sundman linearizibility criteria \cite{mur1}. In the following, we demonstrate the method of deriving ST for this equation. Comparing (\ref{exam2}) with (\ref{main2}), we find $a_2=-\frac{1}{x}$, $a_1=-\frac{1} {t}$ and $a_0=0$. Substituting the latter expressions in (\ref{a2_coeff})-(\ref{jkd}) and solving the resultant equations, we find%get the following expressions
\begin{equation}
r_1=r_2=r_4=0,~~q=0,~~r_3=t,~~f=0.\label{fbgt}
\end{equation}
With these expressions the integral (\ref{k31_new}) read 
\begin{equation}
I=\frac{\dot{x}}{xt}=\frac{\frac{d}{dt}x}{\frac{d}{dt}(\int x t dt)}.\label{eam_int2}
\end{equation}
Eq. (\ref{eam_int2}) gives us
\begin{equation}
w=x,~~z=\int x t dt.
\end{equation}
Substituting (\ref{eam_int2}) in (\ref{sol_sund}) we can obtain the solution of (\ref{exam2}) straightaway in the form
\begin{equation}
x(t)=Ce^{I \frac{t^2}{2}},
\label{eq:}
\end{equation}
where $C$ and $I$ are integration constants.

\subsection{Example 3: Generalized LT}
To demonstrate the method of identifying GLT, we consider a general equation of the form
\begin{equation}
\ddot{x}+3K(t)x^n\dot{x}+K(t)^2x^{2n+1}+\dot{K}(t)x^{2n}+\lambda(t)x=0.\label{main}
\end{equation}
Substituting $n=1$, Eq. (\ref{main}) becomes Eq. (\ref{exam1}). While the latter equation is linearizable through the PT and the former is linearizable by GLT, as we see below.
Comparing the above equation (\ref{main}) with (\ref{main2}) and solving the Eqs.(\ref{a2_coeff})-(\ref{jkd}), we find
\begin{eqnarray}
&q=n+1,\;\;r_{1}=\frac {\ddot{f}} {2\dot{f}},\;\;r_{2}=\frac{3K(t)} {n+2},\;\;r_{3}=\frac {\ddot{f}f} {2\dot{f}}-\frac{9} {(n+2)^2}\dot{f},
\nonumber\\
&r_{4}=\frac{3K(t)f} {n+2},\;\;\lambda=\frac{(n+2)^2} {9}\frac {\dddot{f}} {2\dot{f}}-\frac {3} {4}\frac {\ddot{f}^2} {\dot{f}^2}.\label{para}
\end{eqnarray}
Substituting the above parameters in (\ref{k31_new}) and rearranging it we get the first integral $I$ as
\begin{equation}
I=\frac {\dot{x}+\frac{3K(t)} {n+2}x^{n+1}+\frac {\ddot{f}} {2\dot{f}}x} {f(\dot{x}+\frac {\ddot{f}} {2\dot{f}}x+\frac{3K(t)} {n+2}x^{n+1})-\frac{9}{(n+2)^2}\dot{f}x}.\label{jkfj}
\end{equation}
Now substituting the parameters (\ref{para}) in (\ref{k10}) and we find that the Eq. (\ref{k10}) is not satisfied. Since $f(t) \neq 0$ in the present example it should admit GLT. To obtain the explicit form of it, we substitute the parameters (\ref{para}) in (\ref{k55}). 
Here we find
\begin{equation}
I=\frac{\frac{d} {dt}\bigg(x^{-n}\dot{f}^{\frac{-n} {2}}-\frac{3n} {n+2}\int \dot{f}^{\frac{-n} {2}}K(t)dt\bigg)} {\frac{d}{dt}\bigg(-n \int \dot{f}^{\frac{-n} {2}}(x^{-(n+1)}f\dot{x}+(\frac {\ddot{f}f} {2\dot{f}}-\frac{9} {(n+2)^2}\dot{f})x^{-n}+\frac{3K(t)} {n+2}f)dt \bigg)}.\label{exa_int3}
\end{equation}

Now comparing Eq. (\ref{exa_int3}) with $I=\frac{\frac{dw}{dt}}{\frac{dz}{dt}}$, we find
\begin{eqnarray}
\omega&=&\bigg(\frac{-3n} {n+2}x^{-n}\dot{f}^{\frac{-n} {2}}\int \dot{f}^{\frac{-n} {2}}K(t)dt\bigg),\\
z&=&\bigg(-n \int \dot{f}^{\frac{-n} {2}}(x^{-(n+1)}f\dot{x}+(\frac {\ddot{f}f} {2\dot{f}}-\frac{9} {(n+2)^2}\dot{f})x^{-n}+\frac{3K(t)} {n+2}f)dt \bigg).\nonumber 
\end{eqnarray}

We substitute the parameters given in Eq.(\ref{para}), in Eq.(\ref{jnm_gene}) so that the general solution of Eq. (\ref{main}) can readily be identified as
\begin{eqnarray}
x(t)=\big(Ce^{n\delta}-ne^{n\delta}\int e^{-n\delta}\tilde{f}\frac{3K(t)} {n+2}[1-Ifdt)\big)^{-n},
\end{eqnarray}
where $\delta=\int \tilde{f}\bigg(\frac{\ddot{f}} {2\dot{f}}-I\big(\frac{\ddot{f}} {2\dot{f}}f-\frac{9} {(n+2)^2}\dot{f}\big)\bigg)dt$, $C$ and $I$ are arbitrary constants.

\section{Conclusion}
In this paper, we have developed a method of deriving LTs for a class of second order nonlinear ODEs. Through this procedure one can identify the first integral, LTs and the general solution of the given nonlinear ODE (provided it is linearizable) in a simple and straightforward manner.  We have also demonstrated the proposed algorithm with three different examples. Suppose the given nonlinear ODE is linearizable through multiple LTs the present procedure identifies all of them in a straightforward manner which can be considered as an added feature compare to other existing methods. To derive these linearizing transformations one has to consider a suitable integral.  Interestingly, the same procedure can be extended to identify the LTs in higher order ODEs.  In higher order ODEs it is often cumbersome to derive the LTs.  In those situations the proposed procedure will play a crucial role.  

\section*{Acknowledgments}
RMS acknowledges the University Grants Commission (UGC-RFSMS), Government of India, for providing a Research Fellowship. The work of VKC is supported by INSA young scientist project. The work of MS forms part of a research project sponsored by Department of Science and Technology, Government of India.

\end{document}